\documentstyle[prb,aps,eqsecnum,preprint,fleqn,amssymb]{revtex}

\begin{document}
\draft
\title{Dynamics of a hole in the large--U Hubbard model:\\a Feynman
diagram approach}
\author{C. Zhou and H. J. Schulz}
\address{
Laboratoire de Physique des Solides
\cite{assoc},
 Universit\'{e} Paris-Sud,
91405 Orsay,
France }
\maketitle
\begin{abstract}
We study the dynamics of a single hole in an otherwise  half--filled
two--dimensional Hubbard model by introducing a nonlocal Bogolyubov
transformation in the
antiferromagnetic state. This allows us to rewrite the
Hamiltonian in a form that makes a separation between high--energy processes
(involving double--occupancy) and low--energy physics possible. A
diagrammatic scheme is developped that allows for a systematic study of the
different processes delocalizing a carrier in the antiferromagnetic state.
In particular, the so--called Trugman process, important if transverse spin
fluctuations are neglected, is studied and is shown to be dominated
by the leading vertex corrections. We
analyze the dynamics of a single hole both in the Ising limit and with spin
fluctuations. The results are compared with previous theories as well as
with recent exact small--cluster calculations, and we fing good agreement.
The formalism
establishes a link between weak and strong coupling methodologies.
\end{abstract}

\narrowtext

\section{Introduction}
Strongly correlated fermion systems have been the focus of
interest
in the last few years following the discovery of high temperature
superconductors. At the heart of the problem is the
interplay between the itinerant character of the charge carriers and the
existence of localized spins (on some time scale) imposed by the
single--occupancy constraint. Considerable effort have been invested
in the understanding of the Hubbard model and its strong
coupling cousin, the so--called t--J model, in particular in two dimensions.
At half filling, the strongly correlated Hubbard model reduces to
the Heisenberg model, which is now believed to exhibit
antiferromagnetic order on bipartite lattices in two
dimensions \cite{reger_young}. Further, for arbitrary correlation strength
$U$, the Hartree--Fock solution  leads to
a commensurate spin--density wave (SDW) state
which approaches the N\'eel state as U increases \cite{schrieffer_spinbag}.
In addition, a straightforward RPA calculation for quantum spin fluctuations
has produced  a spin wave mode which in the large $U$ limit coincides
with what is expected from
the Heisenberg model\cite{schrieffer_spinbag,singh_rpa}.

Away from half filling, no consensus exists
yet except the Nagaoka theorem. In the weak--correlation limit
it is known that
doping tends to induce an incommensurate SDW\cite{schulz_gf_incomm}. In
particular, the antiferromagnetic state starts to be discommensurated
at small doping  by deforming into a domain wall
structure\cite{schulz_gf_incomm,poilblanc_hf_incomm,inui_domainwall}.
Increasing the strength of correlations suppresses the amplitude
fluctuations, enhancing instead the instability in the transverse
channel\cite{inui_domainwall}. In particular, a
mean field analysis for the t--J model has predicted a
spiral phase\cite{shraiman_tj_spirale_bis}. However,
inhomogenous phases have also been argued to be stable at least for small
doping even in the large--U limit\cite{arrigoni_incomm,auerbach_inhmphase}.
So far few substantial studies have been devoted to exploit the effects
of quantum fluctuations or to investigate in detail the crossover between
weak and strong correlation in the doped case.

Given the different methodologies used in discussing the weakly and
strongly correlated
regimes of the Hubbard model, it is not at all obvious however why
the smooth interpolation should occur. In particular,
since weak coupling methods are based on the assumption that
the coupling is much less than the bandwidth $8t$, it is
unclear how the local constraint of no double
occupancy is automatically respected
in extrapolating the weak coupling methods to the strong coupling
regime.
The purpose of the present work is partly to fill that gap.
To moderate our ambition we shall
limit ourselves only to the case of a single carrier.
This will allow us to assume the two--sublattice structure
in our analysis, since the antiferromagnetic order is then believed to
persist. Moreover, currently available data on finite
clusters make  reliable quantitative comparisons possible.

The calculations are based on the following idea. By introducing
a nonlocal Bogolyubov rotation
on the Hubbard Hamiltonian  we set up a new basis in which we
can readily see the differences and connections between
weak and strong coupling approaches. In particular, the low energy
scattering processes can be diagrammatically isolated from those at high
energy
without ambiguity. This in turn makes it possible to show how
the latter processes are scaled away as $U/t$ becomes sufficiently large,
thus virtually excluding the possibility of double occupancy in the strong
coupling limit.

The paper is
organized as follows: in section \ref{sec:eff} we introduce the Bogolyubov
rotation
and derive the effective Hamiltonian within the new basis. In
section \ref{sec:half} we take as example the half filled case to examine
quantum spin fluctuations.
The analysis will be performed entirely within the rotated
basis in which (1)
the Hartree--Fock ground state has exactly the same spin configuration as the
unperturbed ``noninteracting'' state, namely the SDW state; (2)
fluctuations above this SDW state can be classified and
diagrammtically distiguished according whether double
occupancies are involved or not. Such a distinction from the
conventional RPA analysis\cite{schrieffer_spinbag,chubukov_rpa,singh_rpa}
proves to be an important advantage.
Based on the
observations made in section  \ref{sec:half}, the Feynman
diagrammatic scheme is then applied to the problem of a single hole
in the strongly correlated regime in section \ref{sec:hole}. We shall
discuss in detail the effects of fluctuations on the
dynamics of a single hole and compare with
results derived using other available techniques.
Comments and a discussion are given in the final section.

\section{The Bogolyubov transformed Hamiltonian}
\label{sec:eff}
The standard Hamiltonian of the Hubbard model is
\begin{equation}
H = -t \sum_{\langle i,j \rangle,\sigma} (C^\dagger_{i,\sigma} C_{j,\sigma} +
h.c.)
+ U \sum_i n_{i,\uparrow} n_{i,\downarrow} \;\;,
\end{equation}
where the $C_{i,\sigma}$ are annihilation operators for a fermion at
site $i$ with spin $\sigma$, $n_{i,\sigma}$ is the corresponding number
operator, and $\langle i,j \rangle$ indicates summation over the nearest
neightbor bonds of a square lattice, each bond being counted once. The ratio
between the onsite
repulsion parameter $U$ and the hopping energy $t$ determines, together
with the bandfilling, the physics of this model.
In this section we introduce the Bogolyubov rotation of the fermion
operators and
then derive an effective Hamiltonian for the strongly correlated case
($U\gg t$)  near half filling.

We begin by going to momentum space and rewriting
the Hamiltonian within the reduced Brillouin zone
(RBZ) defined by $|k_x| + |k_y| \le \pi$. This can be done by
introducing the spinor notion
$\Psi^\dagger_\alpha ({\bbox{k}})=(C^\dagger_{{\bbox{k}}\alpha},
C^\dagger_{{\bbox{k}}+{\bbox{Q}}\alpha})$, where ${\bbox{Q}}=(\pm\pi,\pm\pi)$.
We then obtain (see Appendix A)
\begin{eqnarray}
H &=&\sum_{{\bbox{k}}\alpha}\varepsilon_{\bbox{k}} \Psi^\dagger_\alpha
({\bbox{k}})\sigma^3 \Psi_\alpha ({\bbox{k}})
\nonumber\\
&+
&\frac{U}{N}\sum_{{\bbox{k}}{\bbox{k}}\prime{\bbox{q}}}\{\Psi^\dagger_\uparrow
({\bbox{k}}\!+\!{\bbox{q}})\Psi_\uparrow
({\bbox{k}})\Psi^\dagger_\downarrow ({\bbox{k}}^\prime\!-\!{\bbox{q}})
\Psi_\downarrow
({\bbox{k}}^\prime)
\nonumber \\
& + &[\Psi^\dagger_\uparrow({\bbox{k}}\!+\!{\bbox{q}})\sigma^1\Psi_\uparrow
({\bbox{k}})] [\Psi^\dagger_\downarrow
({\bbox{k}}^\prime -{\bbox{q}})\sigma^1\Psi_\downarrow ({\bbox{k}}^\prime)]\}
\;\;, \label{eq:hred}
\end{eqnarray}
where the $\sigma^i$ ($i=1,2,3$) are the standard Pauli matrices, and
$\varepsilon_{{\bbox{k}}} = -2t(\cos k_x + \cos k_y)$. One should notice that
while the summation over momenta runs only over the
RBZ, ${\bbox{k}}\pm{\bbox{q}}$ can freely go beyond the boundaries of the RBZ.
Each state is then easily seen to be counted once and only once.

Next we set up a new basis by introducing the following
unitary transformation of the fermion operators:
\begin{eqnarray}
\Psi_\sigma ({\bbox{k}})&=&U_\sigma ({\bbox{k}})R_\sigma
({\bbox{k}})\nonumber\\
&=&\left(\begin{array}{rr} u_{\bbox{k}}&v_{\bbox{k}}\\
\sigma v_{\bbox{k}}&-\sigma
u_{\bbox{k}}\end{array}\right)\left(\begin{array}{l}
r^c_{{\bbox{k}}\sigma}\\r^v_{{\bbox{k}}\sigma}\end{array}\right) \label{eq:su2}
\label{eq:rotation} \;\;,
\end{eqnarray}
and for $u_{\bbox{k}}, v_{\bbox{k}}$ we take the large--$U$ expansion
\begin{equation}
u_{\bbox{k}}^2=\frac{1}{2}(1+\frac{2\varepsilon_{\bbox{k}}}{U})
\quad \quad \;\;, \quad \quad
v_{\bbox{k}}^2=\frac{1}{2}(1-\frac{2\varepsilon_{\bbox{k}}}{U})
\;\;.
\end{equation}
This of course is the Bogolyubov transformation used
in the weak coupling approach in diagonalizing
the Hartree--Fock Hamiltonian \cite{schrieffer_spinbag}.
However, we emphasize that the operation here will  be applied
to the full Hamiltonian instead of the mean field one, and therefore
is exact irrespective of the form
of the coefficients $u_{\bbox{k}},v_{\bbox{k}}$.
As it will soon be clear this procedure will
permit us to examine to what extent the weak coupling
approach overlaps with the strong coupling methods.

The transformed Hamiltonian is
rather involved when written in momentum space and includes both
interband and intraband
scatterings as well as conditional single particle interband hoppings.
To render things more transparant we turn instead to real space.
Assuming $U/t\gg 1$ and keeping only terms up to O($t^2/U^2$),
we can easily rewrite eq.(\ref{eq:rotation}) in real space and obtain
a nonlocal transformation: for the site $i$ on the even sublattice
and $\sigma = \downarrow$ or site $i$ on the odd sublattice and
$\sigma = \uparrow$ one has
\begin{equation}
\label{eq:tr1}
C_{i\sigma}=r^v_{i}-
\frac{t}{U}\sum_{\delta}r^c_{i+\delta}-
\frac{J}{8U}\sum_{\delta ,\delta^\prime}r^v_{i+\delta-
\delta^\prime} \;\;,
\end{equation}
whereas in the opposite case
\begin{equation}
\label{eq:tr2}
C_{i\sigma}=r^c_{i}+
\frac{t}{U}\sum_{\delta}r^v_{i+\delta}-
\frac{J}{8U}\sum_{\delta ,\delta^\prime}r^c_{i+\delta-
\delta^\prime} \;\;.
\end{equation}
and $J=4t^2/U$ as usual.
The main difference between the new basis
and the original one is thus very  clear:
the electrons are now classified in terms of the
category ($c$ or $v$) they belong to, instead of spin:
a $v$ electron has spin $\downarrow$ if it sits on the even sublattice
and spin $\uparrow$ if it sits on the odd sublattice, and vice versa for
the $c$--electrons. Thus,
a fully occupied $v$--band is essentially a N\'eel
ordering in the original basis (namely, we observe a downspin
on the even sublattice and an upspin  on the odd sublattice).
We note that the particles created by the $r$--operators are not strictly
localized on their respective sites, but are rather in the Wannier states
corresponding to the mean--field N\'eel state and thus are to a certain degree
delocalized on nearest (amplitude $\propto t/U$) and next--nearest
(amplitude $\propto (t/U)^2$) neighbors.

The effective Hamiltonian is finally obtained by substituting the above
tranformation into the Hubbard Hamiltonian. We thus have, retaining
only terms up to $O(t^2/U)$:
\begin{equation}
H=H_1+H_2+H_d \;\;,
\label{eq:Heff}
\end{equation}
\begin{eqnarray}
H_1 =&-&t\sum_{i\delta}\{(1-n^v_i)
r^{c \dagger}_ir^v_{i+\delta}(1-n^c_{i+\delta})+h.c.\}\nonumber\\
&+&2J\sum_i(n^c_i-n^v_i) - J\sum_{\langle
i,j\rangle}(S^z_iS^z_j-\frac{1}{4}n_in_j)
+\frac{J}{2}\sum_{\langle i,j\rangle}(S^\dagger_iS^-_j+S^-_iS^\dagger_j)
\;\;, \label{eq:H1}
\end{eqnarray}
\begin{eqnarray}
H_2 =
& &\frac{J}{2}\sum_{i,\delta\ne\delta^\prime}(r_i^{c \dagger}
r^c_{i+\delta-\delta^\prime}-
r_i^{v \dagger}r^v_{i+\delta-\delta^\prime})
+\frac{J}{4}\sum_{i,\delta\ne\delta^\prime}
(n^c_ir^{c \dagger}_{i+\delta}r^c_{i+\delta\prime}+
n^v_ir^{v \dagger}_{i+\delta}r^v_{i+\delta\prime}) \nonumber\\
&-&\frac{J}{4}\sum_{i,\delta\ne\delta^\prime}
(r_i^{c \dagger}r^v_{i+\delta}r^{c \dagger}_{i+\delta^\prime}r_i^v
+h.c.) \;\;,
\label{eq:H2}
\end{eqnarray}
\begin{eqnarray}
H_d =
& &U\sum_in^c_in^v_i
-\frac{J}{8}\sum_{i,\delta\delta^\prime}
\{n^c_i(r_i^{v \dagger}r_{i+\delta-\delta^\prime}^v+h.c.)
+c\leftrightarrow v\}\nonumber\\
&-&\frac{J}{4}\sum_{i,\delta\delta^\prime}(r_i^{c \dagger}r^v_{i+\delta}
r_i^{v \dagger}r^c_{i+\delta^\prime}+h.c.)
+t\sum_{i\delta}(n^c_ir^{v \dagger}_ir^c_{i+\delta}n^v_{i+\delta}+h.c.)
\;\;. \label{eq:Hd}
\end{eqnarray}
The spin index in the original Hamiltonian now is replaced by the $v$
and $c$ band indices, as discussed following eqs. (\ref{eq:tr1}) and
(\ref{eq:tr2}). The spin operators  in $H_1$
are defined by $S^\dagger_{i}=r^{c \dagger}_{i}r^v_{i}$ on the even
sublattice
and $S^-_{i}=r^{\dagger
c}_{i}r^v_{i}$ on the odd sublattice and
represent local spinflips in the {\it new} basis. Further,
$n_in_j/4-S^z_iS^z_j$ in $H_1$ is a rewriting of
$(n^c_in^c_j+n^v_in^v_j)/2$ which represents nearest neighbor
intraband scattering. We have expressed the Hamiltonian in
three terms, namely $H_1$, $H_2$, $H_d$, to emphasize the
fact that all interaction processes in $H_d$ involve double occupancy,
and in $H_2$ the electrons will always go beyond the nearest
neighbors. We also notice that in $H_1$ and $H_d$ there are terms
involving six fermionic operators. These terms in fact cancel (not
surprisingly, given that the original Hamiltonian only involves terms
containing up to four fermion operators and that the transformation to
the new operators is linear), however, we find the forms (\ref{eq:H1}),
(\ref{eq:Hd}) rather useful because they clearly separate terms
involving double occupancy from those which don't.

It is clear that at half--filling (one electron per site) the
``non--interacting'' (or Hartree--Fock) ground state of the tranformed
Hamiltonian (\ref{eq:Heff}) is not the
paramagnetic Fermi sea as in the original basis, but rather a state with
all the $v$--states filled and all the $c$--states are empty. That is to
say that in the new basis electrons
will first fill the $v$--band in which the spin of each electron is seen to
point down on the even sublattice and up on the odd sublattice.
The Hartree--Fock energy of this state is $E_0=-NJ$.

The Hartree--Fock quasiparticle energies are obtained by calculating the
average value of $H$ in a state with either a $c$--particle added or a
$v$--particle taken out. One then finds
\begin{equation}
H_{HF}=\sum_{{\bbox{k}}}
( \xi^c_{\bbox{k}} r^{c\dagger}_{{\bbox{k}}}r^c_{{\bbox{k}}}
+\xi^v_{\bbox{k}} r^{v \dagger}_{{\bbox{k}}}r^v_{{\bbox{k}}} ) -NJ \;\;,
\label{eq:HF1}
\end{equation}
where $\xi^c_{\bbox{k}}=U-J+\varepsilon^2_{{\bbox{k}}}/U$ and
$\xi^v_{\bbox{k}}=J-\varepsilon^2_{{\bbox{k}}}/U$, and the $k$--space operators
are
defined by
\begin{equation}
r^{c,v}_{{\bbox{k}}} = \frac{1}{\sqrt{N}} \sum_i r^{c,v}_i e^{i {\bbox{k}}
\cdot
\bbox{R}_i} \;\;,
\end{equation}
with the $k$--summation covering the full Brillouin zone. We note that the
operators thus defined are {\em not} eigenfunctions of $\sigma_z$, because
the spin projection of e.g. a $v$--particle is different on the even and
odd sublattice. Spin eigenstates can however be straightforwardly
constructed noting that $\xi^{c,v}_{{\bbox{k}}} =
\xi^{c,v}_{{\bbox{k}}+\bbox{Q}}$. The operators with well-defined spin then
are $r^{c,v}_{{\bbox{k}}\sigma} = (r^{c,v}_{{\bbox{k}}} \pm
r^{c,v}_{{\bbox{k}}+\bbox{Q}})/\sqrt2$, where $k$--space is now limited to
the {\em reduced} Brillouin zone. In terms of these operators the
Hartree--Fock Hamiltonian takes the form
\begin{equation}
H_{HF}=\sum_{{\bbox{k}}\sigma}
( \xi^c_{\bbox{k}} r^{c \dagger}_{{\bbox{k}}\sigma}r^c_{{\bbox{k}}\sigma}
+\xi^v_{\bbox{k}} r^{v \dagger}_{{\bbox{k}}\sigma}r^v_{{\bbox{k}}\sigma} ) -NJ
\;\;. \label{eq:HF}
\end{equation}
The $c$--band is now seen to be separated from the fully occupied $v$--band
by a gap of order $U$. In addition, the Hartree--Fock ground state at
half--filling has a fully filled $v$--band  and an empty
$c$--band corresponding to N\'eel spin ordering.

Going back to the full Hamiltonian (\ref{eq:Heff}) it is clear that
for electron density
$n\le 1$ the $c$--band is drawn in only through {\it interband}
interactions. To get an idea how the $c$--band becomes involved,
we notice that
interband interactions can actually be divided into two classes.
Interband processes described in $H_d$ can bring in the $c$--band but
always provoke double occupancy simultaneously. At the Hartree--Fock level
this of course pushes the $c$--band by $U$ above the $v$--band, and  it is
clear that beyond Hartree--Fock, there will always an intermediate state of
energy $\approx U$ involved. On the other hand,
interband processes contained in $H_1+H_2$ can actually transfer the
$v$--band
electrons to the $c$--band or vice versa at much less energy expense. One
such example is the interchange of two nearest neighboring $v$--band
particles.
This amounts to having a spinflip at each site since these two
particles have opposite spins, or equivalently the pair of
$v$--band particles hops to the $c$--band in the end. Such a
process is realized via the interband pair hoping described by
the last term in $H_1$ and costs only an energy of order $O(J)$.

It is physically clear that scatterings involving double occupancy
will actually be circumvented in the strong
coupling limit and therefore render no substantial
contribution to the low energy physics.  Nevertheless, this is not so
obvious
if we deal directly with the Hubbard model using Feynman's diagrammatic
technique. It turns out and will be shown below that the advantage of
the Bogolyubov rotation proposed here is that it provides a basis
in which we are able to clearly demonstrate with Feynman diagrams the
irrelevence of $H_d$ in the effective low energy properties.

Once established that $H_d$ will eventually be scaled away,
we are left with $H_1+H_2$
to describe low energy physics of the large--$U$ Hubbard
model. In particular, $H_1$  is scarcely different from the
familiar $t-J$ model, except for the interactions in the (charge and
spin) density channel.
Here we note that the ``noninteracting'' state we start
with in the new basis exhibits N\'eel ordering. The spin
fluctuations above the N\'eel state are then of ferromagnetic
nature in the longitudinal spin channel. On the
other hand, since the spin rotational symmetry has been retained in the
transformation fluctuations in the transverse spin channel are
described in exactly the same manner as in the t--J model.

\section{Quantum fluctuations at half filling}
\label{sec:half}
We devote this section to the discussion of
quantum spin fluctuations in the half--filled Hubbard
model. The final result is actually known either from spin wave
calculations or from direct extrapolation of
weak--coupling results. The convergence of these apparently unrelated
approaches raises the question of a possible
smooth crossover between the two extreme coupling regimes.
The main point which remains unclear is
why and how the local constraint of no double occupancy becomes
automatically respected when one goes to large $U$ within the
mean--field formalism which is frequently believed to be limited to  weak
coupling.
The purpose of our presentation is to demonstrate that within the new
basis we are able to show explicitly and in Feynman
diagrams how the higher
energy processes involving double occupancies
({\it i.e.} $H_d$)
are in the end scaled away and become irrelevant to the spin dynamics.

The antiferromagnetic ordering implies the the existence
of gapless spin wave excitations in the transverse spin channel.
This can be analyzed by studying the
dynamical spin susceptibility in the
presence of N\'eel
ordering which is now defined as a $2\times 2$ matrix:
\begin{equation}
\chi^{-+}({\bf x},{\bf x}^\prime,t)=
-i \left(\begin{array}{rr}
\langle
T\left(\bar{S}^-_{x_e}(t)\bar{S}^+_{x^\prime_e}(0)\right)\rangle &
\langle
T\left(\bar{S}^-_{x_e}(t)\bar{S}^+_{x^\prime_o}(0)\right)\rangle \\
\langle
T\left(\bar{S}^-_{x_o}(t)\bar{S}^+_{x^\prime_e}(0)\right)\rangle &
\langle
T\left(\bar{S}^-_{x_o}(t)\bar{S}^+_{x^\prime_o}(0)\right)\rangle
\end{array}\right) \;\;.
\label{eq:matrixA}
\end{equation}
As convention, the position coordinate ${\bf x}$ is denoted by ${\bf
x_e}$ if the spin is on the even
sublattice and by ${\bf x_o}$ if on the odd sublattice.
We have added a bar over the spin operators in order to
emphasize that we are probing
correlations between spins defined in terms of the original $C$--operators
(as opposed to the $S$--operators occurring in eq.(\ref{eq:H1})).

The calculation for the above matrix at the Hartree--Fock level is
straightforward. We first transform the susceptibility matrix such
that $\bar{S}^+$, $\bar{S}^-$ are written in terms of $r$--operators
as presented in Appendix B. The leading contribution is then
readily obtained. Up to $O(J/U)$ only the diagonal terms in
eq.(\ref{eq:matrixA}) are nonzero and are given by
\begin{equation}
\chi^{-+}_{HF11}({\bbox{q}},\omega)=
\chi^{-+}_{HF22}({\bbox{q}},-\omega)=
\frac{2}{N}\sum_{\bbox{k}}
\frac{1}{\omega-\xi^c_{{\bbox{k}}-{\bbox{q}}}+\xi^v_{\bbox{k}}+i0^+}\;\;.
\label{eq:HFbubble}
\end{equation}
With little algebra\cite{footnotes2} one can verify that
this agrees with the results\cite{schrieffer_spinbag} of SWZ up to
$O(J/U)$.

To calculate the dynamical susceptibility in the presence of
interactions we need to take some care.  In all previous
studies\cite{schrieffer_spinbag,singh_rpa,chubukov_rpa},
the RPA scheme was adopted within the
{\it paramagnetic} $C$--operator basis in which the
interaction in the transverse spin channel is exactly
$U$. Here we rather write the original spin operators ($\bar{S}$) in terms
of the $r$--operators. Then the simplest
Hartree--Fock ground state already has the
correct antiferromagnetic structure, and the interaction between the
Hartree--Fock particles is directly obtained from
eqs.(\ref{eq:Heff})--(\ref{eq:Hd}).
 To facilitate
the following  discussion we write the interaction explicitely in
momentum space. The terms relevant for the transverse spin fluctuations
(\ref{eq:matrixA}) are
\begin{eqnarray}
H^{+-}_I &=&
\frac{2}{N}\sum_{{\bbox{k}}{\bbox{k}}\prime{\bbox{q}}} \left(
g_1({\bbox{k}},{\bbox{k}}^\prime)
r^{c \dagger}_{{\bbox{k}}+{\bbox{q}}\downarrow}
r^{v \dagger}_{{\bbox{k}}\prime-{\bbox{q}}\uparrow}r^v_{{\bbox{k}}\uparrow}
r^c_{{\bbox{k}}^\prime\downarrow}
+ g_2({\bbox{k}},{\bbox{k}}^\prime)
r^{c \dagger}_{{\bbox{k}}\downarrow}
r^{c \dagger}_{{\bbox{k}}\prime-{\bbox{q}}\uparrow}
r^v_{{\bbox{k}}-{\bbox{q}}\uparrow} r^v_{{\bbox{k}}\prime\downarrow}
\right. \nonumber \\
& & \left. +v\longleftrightarrow c\right) \;\;,
\label{eq:fluc}
\end{eqnarray}
with
\begin{equation}
g_1({\bbox{k}},{\bbox{k}}^\prime)=
U\left[ 1-\frac{\varepsilon^2_{\bbox{k}}+
\varepsilon^2_{{\bbox{k}}+{\bbox{q}}}+\varepsilon^2_{{\bbox{k}}^\prime}
+\varepsilon^2_{{\bbox{k}}^\prime-{\bbox{q}}} }{2U^2} \right] \;\;,
\label{eq:g1}
\end{equation}
\begin{equation}
g_2({\bbox{k}},{\bbox{k}}^\prime)=
-\frac{\varepsilon_{\bbox{k}}\varepsilon_{{\bbox{k}}-{\bbox{q}}}
+\varepsilon_{{\bbox{k}}\prime}\varepsilon_{{\bbox{k}}\prime-{\bbox{q}}}}{U}
\;\;. \label{eq:g2}
\end{equation}
We note that no $O(t)$ terms enter as can easily be understood for half
filling.
Neither do the intraband interactions enter since they contribute only to
longitudinal fluctuations. $H^{+-}_I$ is now composed of two parts.
The $g_1$ part is equivalent to the first two terms in
$H_d$ and describes interband scatterings involving double occupancy.
On the other hand, the $g_2$ part
describes the transfer of a pair of particles between two bands and
corrsponds to the last terms in $H_1$ and $H_2$.
This part will be shown to be the one responsible for the spin dynamics.
Finally it will be helpful to note that up to $O(J/U)$,
$g_1$ and $g_2$ can actually be factorized
by $g_1=Uf_1({\bbox{k}})f_1({\bbox{k}}^\prime)/4$ and
$g_2=U(f_1({\bbox{k}})f_2({\bbox{k}}^\prime)
+f_1({\bbox{k}}^\prime)f_2({\bbox{k}}))/4$,
with $f_1$ and $f_2$ being defined in appendix B.

We are now in the position to study the transverse spin fluctuations
in the SDW state. Following appendix B we shall calculate
\begin{eqnarray}
\chi^{-+}(\bbox{q},\omega )=-\frac{i}{2N}\sum_{{\bbox{k}}{\bbox{k}}^\prime}
f_1({\bbox{k}})f_1({\bbox{k}}^\prime)
\left(\begin{array}{cc}a_{11} &a_{12} \\
a_{21} & a_{22} \end{array}\right)
\label{eq:matrix}
\end{eqnarray}
to get the dominant contribution. The definition
for the matrix elements is given in appendix B.
A slight complication arises in the calculation due to
the fact that in the interaction Hamiltonian $H^{+-}$
two interaction vertices $g_1$ and $g_2$ are involved, and
both of them are momentum--dependent.
However, we should notice that the $g_1$ interaction
conserves the number of particles in each band. In other words, for an
excited particle--hole pair of opposite spin the $g_1$--term
contributes to
multiple scatterings between the particle and hole.
On the contrary $g_2$--interactions give rise to
destruction and formation of p--h pairs of different spins.
Within the RPA scheme we then find that $g_1$ and $g_2$ will operate
distinctly and the two interactions do not mix up
in Feynman diagrams. A typical diagram for $\chi^{-+}$ incorporating
both interactions is shown in fig.\ref{fig:rpa}.

Having learnt that the $g_1$-- and $g_2$--type interactions do not mix
up at RPA level, we can deal with them separately. Let us
look at the $g_1$--interaction first.
As is evident, the off--diagonal
terms in eq.(\ref{eq:matrix}) remains zero with only $g_1$
interactions. To get the diagonal terms of eq.(\ref{eq:matrix}), we need to
evaluate the diagram illustrated in fig.\ref{fig:rpa}b.
After a straightforward calculation we obtain
\begin{equation}
\chi^{-+}_{11(22)} (\bbox{q},\omega ) =\frac{\frac{1}{2N}\sum_{\bbox{k}}
f^2_1
({\bbox{k}}) A({\bbox{k}},{\bbox{q}})}
{ 1+\frac{U}{2N}\sum_{\bbox{k}} f^2_1 ({\bbox{k}}) A({\bbox{k}},{\bbox{q}}) }
\;\;, \label{eq:g1renor}
\end{equation}
where $1/A({\bbox{k}},{\bbox{q}})=\pm\omega-
\xi^c_{{\bbox{k}}-{\bbox{q}}}+\xi^v_{\bbox{k}}+i0^+ $.
In particular, taking the limit of $U\gg t$ gives
the $g_1$--renormalized susceptibility as
\begin{equation}
\chi^{-+}_0({\bbox{q}},\omega)=\left(
\begin{array}{cc}
\chi^0({\bbox{q}},\omega)&0\\0&\chi^0({\bbox{q}},-\omega)
\end{array} \right) \;\;,
\label{eq:chi0}
\end{equation}
\begin{equation}
\chi^0({\bbox{q}},\omega)=\frac{1}{\omega-2J+i0^+} \;\;.
\label{eq:bubble}
\end{equation}
We notice that going
from eq.(\ref{eq:HFbubble}) to eqs.(\ref{eq:chi0}), (\ref{eq:bubble})
the high energy scale of
$O(U)$ has disappeared due to the renormalizaion by multiple $g_1$
scatterings and has been replaced by $2J$ which is
the energy one needs to create a local spin flip. This is natural because
after all a local spin flip can occur without involving double occupancies.

Finally we switch on the $g_2$ interactions.
As explained before and shown in fig.\ref{fig:rpa} within RPA
they have no effect on the
structure within the bubble but play their role by forming junctions between
$g_1$--renormalized bubbles.
Besides, $g_2$ interations also render a nonzero contribution to
off--diagonal
terms $a_{12}$ and $a_{21}$ in eq.(\ref{eq:matrix}), as is shown
diagrammatically in fig.\ref{fig:rpa2}.
Again the work involved in the calculation is routine since
$g_2$ also factorizes as $g_2({\bbox{k}},{\bbox{k}}^\prime)\sim
(f_1({\bbox{k}})f_2({\bbox{k}}^\prime)+f_1({\bbox{k}}^\prime)f_2({\bbox{k}}))$.
We only mention that because of the way in which $g_2$ is factorized, we
shall meet in the intermediate states bubble diagrams
like fig.\ref{fig:rpa}b but with vertex parts $f_1$
or $f_2$ attached to one of its two legs.
For example, when $f_1$ is attached to one leg and $f_2$ to
the other, the diagram gives $4J\gamma_{\bbox{q}}/(U(\pm\omega-2J))$, whereas
$f_2$ is attached to both legs leads to
$4J^2\gamma^2_{\bbox{q}}/(U^2(\pm\omega-2J))$.
The final result for the susceptibility then is obtained as
\begin{equation}
\chi^{-+}({\bbox{q}},\omega )=-\frac{2J}{\omega_{\bbox{q}}^2-\omega^2}
\left(\begin{array}{rr}
1+\frac{\omega}{2J}&-r_{\bbox{q}}\\
-r_{\bbox{q}}&1-\frac{\omega}{2J}
\end{array}\right) \;\;,
\label{eq:mode}
\end{equation}
where $\omega_{\bbox{q}}=2J\sqrt{1-r_{\bbox{q}}^2}$ is the
standard spin wave spectrum. The same expressions can be found from
Singh and Tesanovic's results\cite{singh_rpa,footnotes3} by
directly extrapolating
the RPA results.  However, with the help of Bogolyubov
rotation and by working within the rotated basis we have explicitly
shown here how the higher energy scale
gets irrelevant in the spin dynamics within the framework of a diagrammatic
approach.

At last, we point out that the above results can equally be derived
by considering only $H_1$ and neglecting $H_2$ and $H_d$ entirely,
using the same Feynman diagrammatical
scheme. From the Hartree--Fock solution of $H_1$ we immediately get
eq.(\ref{eq:chi0}), and the final result of eq.(\ref{eq:mode}) is
obtained following RPA for the Heisenberg exchange interactions in the
transverse spin channel.
The coincidence teaches us two things. First, it shows that the
$t-J$ model represents the Hubbard model more faithfully  in the
strongly correlated regime than the $t-J-J^\prime$ or $t-t^\prime-J$ models.
Secondly, one sees that it is possible to extend the Feynman
diagrammatic technique to the strong coupling regime by working
directly within the sub--Hilbert space of no double occupancy.
In the next two sections we shall use this fact to discuss the
dynamics of a single hole in the otherwise quantum antiferromagnetic
background.

\section{The dynamics of a single hole}
\label{sec:hole}
We now begin to examine the motion of a hole in the
antiferromagnetic environment. For this purpose we
take away one downspin electron  from the $v$--band and
study the Green's function for the $v$--band hole
\begin{equation}
{\cal G}^v_\sigma ({\bbox{k}},\omega) =\langle 0|r^{v
\dagger}_{{\bbox{k}}\sigma}\frac
{1}{\omega-H+i0^+}r^v_{{\bbox{k}}\sigma}|0\rangle
\label{eq:green}
\end{equation}
within the sub--Hilbert space of no double occupancy, as defined in
the last section. Here $|0\rangle$ represents the full $v$--band, i.e.
the N\'eel state.
For the reason given at the end of the last section,
and also to render possible a direct comparison
with available  numerical results obtained notably for the
$t-J(J_z)$ model
we shall in the following (and unless specified otherwise) neglect
$H_2$ which always leads to jumps beyond the nearest
neighbors. The principal
subject we shall deal with is therefore $H_1$.

To begin we shall also neglect spin fluctuations altogether ($t-J_z$
model). The Hartree--Fock state of $H_1$ is defined
by two degenerate energy levels ($\omega_v=0$ and
$\omega_c=2J$ respectively). By ignoring spin fluctuations
the only remaining interaction  in $H_1$ is
the $O(t)$ hopping term which takes effect when a hole is introduced and
is allowed to move:
\begin{equation}
H_I = - t\sum_{i\delta}\{
(1-n^v_i) r^{c \dagger}_i r^v_{i+\delta}(1-n^c_{i+\delta})
+(1-n^c_{i+\delta}) r^{v \dagger}_{i+\delta}r^{c}_i (1-n^v_i)
\} \;\;.
\label{eq:HI}
\end{equation}
Obviously, this term correlates the hopping motion of a particle (or
hole) with the occupation of its surrounding sites and thus is
responsible for the coupling between the moving hole and
its N\'eel environment. We now notice that $H_I$ only destroys
particles on singly occupied sites (double occupancy of a site means
that there are both a $c$-- and a $v$--particle) and only creates
particles on empty sites. Therefore acting on a state containig only
empty and singly occupied sites as in (\ref{eq:green}) one can
rewrite $H_I$ in the simpler form
\begin{equation}
H_I = - t\sum_{i\delta}\{
(1-n^v_i) r^{c \dagger}_i r^v_{i+\delta}
+r^{v \dagger}_{i+\delta}r^{c}_i(1-n^v_i)
\} \;\;,
\label{eq:HI2}
\end{equation}
which now contains four--fermion terms only. For later
convenience we write down its Fourier transform in the RBZ explicitely:
\begin{equation}
H_I=-\frac{2}{N}\sum_{{\bbox{k}}{\bbox{k}}^\prime{\bbox{q}}}\{
\varepsilon_{{\bbox{k}}^\prime}\tilde{r}^{v \dagger}_{{\bbox{k}}\downarrow}
\tilde{r}^v
_{{\bbox{k}}-{\bbox{q}}\downarrow}\tilde{r}^{c
\dagger}_{{\bbox{k}}^\prime-{\bbox{q}}\uparrow}
\tilde{r}^v_{{\bbox{k}}^\prime\uparrow}
+\varepsilon_{\bbox{k}}\tilde{r}^{v \dagger}_{{\bbox{k}}^\prime\uparrow}
\tilde{r}^v_{{\bbox{k}}^\prime+{\bbox{q}}\uparrow}
\tilde{r}^{v \dagger}_{{\bbox{k}}\downarrow}\tilde{r}^c
_{{\bbox{k}}-{\bbox{q}}\downarrow}+h.c.\} \label{eq:inter} \;\;,
\end{equation}
where
$\varepsilon_{\bbox{k}}=-zt\gamma_{\bbox{k}}
=-2t(\cos({\bbox{k}}_x)+\cos({\bbox{k}}_y))$.
We note in particular that the momentum in  $\varepsilon_{\bbox{k}}$ is
always that
of the the $v$--band hole which has the same spin as the $c$--particle.
This fact will be used on numerous occasions when we need to determine
vertex functions. In addition, the hole notation $\tilde{r}=r^\dagger$
has been introduced in order to facilitate the discussion.

\subsection{The incoherent spectrum}
It is evident that the first nonzero contribution arises
only in  the second order correction to the selfenergy.
Naively one would have expected, as is readily derived from the
interaction
Hamiltonian, contributions from the two diagrams shown in
fig.\ref{fig:2nd}. Nevertheless, in the absence of spin fluctuations
the contribution from fig.\ref{fig:2nd}b vanishes since a
the lower ($v,\uparrow$) line  represents hole propagation, which implies
that the ($c,\downarrow$) line should represent creation of a hole in the
$c$ band. This of course is impossible, the $c$ band being initially empty.
In other words,
a pair of a {\it $c$--particle} and a {\it $v$--hole}
of opposite spins can not appear {\it before} the hole starts moving.
Consequently for an initial downspin hole we only see
in the intermediate state an upspin hole, accompanied by a
p--h pair formed by $c$--band upspin particle and $v$--band downspin
hole. The diagram is then evaluated according to the
standard Feynman rules and is given by:
\begin{equation}
\Sigma^v_{\downarrow}({\bbox{k}},\omega+i0^+ )
=i\frac{2z^2t^2}{N}\sum_{{\bbox{k}}_1}
\gamma^2_{{\bbox{k}}_1}\int\frac{d\omega_1}{2\pi}
{\cal G}^{v(0)}_{\uparrow}({\bbox{k}}_1,\omega_1+i0^+)
\chi^0(\bbox{k}-{\bbox{k}}_1, \omega-\omega_1) \;\;.
\label{eq:2nd}
\end{equation}
The hole notation has been emphasized in the expression by
$\omega+i0^+$
such that the Green function of a hole below the Fermi sea is seen as
a
retarded function of time. The noninteracting Green function is therefore
${\cal G}^{v(0)}=1/(\omega+i0^+)$, the p--h bubble is easily
determined and is given by
eq.(\ref{eq:bubble}) which is also a retarded function. Moreover, the
vertex functions at two intertacting lines are all determined
by the momentum ${\bbox{k}}_1$ of the scattered hole in the
intermediate state. We thus obtain as final result:
\begin{equation}
\Sigma^v_{\downarrow}({\bbox{k}},\omega)
=\frac{zt^2}{\omega-2J} \;\;.
\end{equation}
The factor z reflects that the hole moves to its z nearest neigbors.
The absence
of momentum dependence corresponds to the fact that
the approximation actually binds holes to its origin by allowing
only back and forth motions towards its nearest neighbors (as implied
by the momentum summation over $\gamma^2_{\bbox{k}}$). Along this line,
the p--h pair in fig.\ref{fig:2nd}a is easily seen to represent
the local spinflip the hole creates at its initial position once it hops.

A more complete picture arrises when we introduce selfconsistency in
fig.\ref{fig:2nd}a. The procedure is immediately seen to permit the
intermediate upspin hole to move further forward.
The new Feynman diagram (fig.\ref{fig:self2nd}) requires us to
solve the following selfconsistency equation:
\begin{equation}
\Sigma^v_{\sigma}({\bbox{k}},\omega+i0^+ )
=i\frac{2z^2t^2}{N}\sum_{{\bbox{k}}_1}\gamma^2_{{\bbox{k}}_1}
\int\frac{d\omega_1}{2\pi}
{\cal G}^v_{-\sigma}({\bbox{k}}_1,\omega_1+i0^+)\chi^0({\bbox{k}}-{\bbox{k}}_1,
\omega-\omega_1) \;\;.
\label{eq:se1}
\end{equation}
After substituting the bare p--h bubble, eq.(\ref{eq:bubble}),
the coupled equations for $\Sigma$ and $\cal{G}$ are further reduced to
a simple iterative equation for the spin and momentum independent Green
function:
\begin{equation}
{\cal G}^v(\omega )=\frac{1}{\omega -zt^2{\cal G}^v(\omega-2J)}
\;\;. \label{eq:string}
\end{equation}
This agrees with the
result of Shraiman and Siggia for the incoherent hole
spectrum\cite{shraimansiggia_string_1hole,martinez_slave_1hole}.

The physics is rather clearly reflected in the diagrammatic picture.
Because of the selfconsistency, the selfenergy is now described by a series
of Feynman diagrams of order $2,4,6,...$. At a
typical order $2n$ (fig.\ref{fig:SS}), a sequence of $n$ bubbles
is first formed and the bubbles are then eliminated one after another in
{\it reversed} order.
Since a momentum--independent bubble of opposite spins in the Feynman
diagrams
means a local spinflip in real space,  the above sequence describes
actually a string of overturned spins in the intermediate state. In
addition,
the number of bubbles $n$ is the length of the string for that
particular order. In conclusion the selfconsistency
has, on the one hand, made possible to
allow for infinitely many particle--hole excitations
in the intermediate states along the way the hole hops forward; on the
other hand it guarantees that the hole can only follow the
self--retracable
paths defined by Brinkman and Rice\cite{brinkmanrice_1hole} and
remains bound to its original position. We have thus properly
incorporated
the contributions from incoherent excitations within the
diagrammatic scheme.

The numerical solution of eq(\ref{eq:string}) defines a ladder
spectrum or a sequence of bound states, with the lowest state
being order of $t(J/t)^{2/3}$ above the lower edge of the
Brinkman--Rice band $\omega_0=-2\sqrt{z}t$.
The discrepancy with the original
$-2\sqrt{z-1}t$ of Brinkman and Rice\cite{brinkmanrice_1hole} is understood
according to Kane et al\cite{kane_slave_1hole}:
in summing $\gamma^2_{\bbox{k}}$ in eq.(\ref{eq:se1})
contributions from all the nearest neighbors are assumed at every stage of
the intermediate states. This results
in overcounting from the second step on,  since in fact only
forward going steps should be counted\cite{martinez_slave_1hole}.
Finally we mention that inclusion of $H_2$ in the above
calculation would have destroyed the ladder structure because then the
hole is already mobile at the Hartree--Fock level. Strictly speaking,
this term is of course present (and of order $t^2/U$) in the Hubbard model
(but not in the $t-J$ model).

\subsection{The coherent spectrum: Trugman's motion}
So far the hole is completely localized. In the absence of spin
fluctuations
it becomes mobile only when it manages to move without disturbing the
antiferromagnetic background.
As Trugman has observed\cite{trugman_loop_1hole}, the
simplest path to permit this is one where the hole hops around
a plaquette one and a half times. The hole then moves to
its nearest neighbor on the same sublattice
while cleaning up the string of reversed spins in the end.
In the diagrammatic scheme, we need
to resort to vertex corrections to account for such paths.
In fact, earlier studies  have realized that
Trugman's  path corresponds to the leading nonzero vertex correction
involving two loops\cite{martinez_slave_1hole,johnson_loop_1hole}.
Within the present framework,
the same process is described by a unique diagram as shown
in fig.\ref{fig:vertex6}.
The calculation for this bare 6th order vertex correction is
straightforward noting that
both the hole lines and
the particle--hole bubbles in the diagram are momentum independent and
{\it retarded}  in time. A slight complication occurs only  due to the
momentum dependence in the interaction vertex.
For our purposes we only need to keep in mind that
the vertex function at each interaction line in the diagram
is  determined by the momentum of
the scattered hole in the intermediate state (the bold line in
fig.\ref{fig:self2nd}, as already encountered in obtaining
eq.(\ref{eq:2nd}). The rules for calculating the diagrams are again the
standard Feynman
rules. The result for the bare 6th order correction is then found to be
\begin{equation}
\Sigma^{v(6)}_\downarrow ({\bbox{k}},\omega+i0^+ )
=\frac{z^4t^6}{(\omega-2J)^2(\omega-4J)^2(\omega-6J)}\frac{2}{N}\sum_{\bbox{q}}
\gamma^2_{\bbox{q}}\gamma^2_{{\bbox{k}}+{\bbox{q}}} \;\;.
\label{eq:vertex6}
\end{equation}
The denominator in the above expression can easily be explained
following fig.\ref{fig:vertex6}b. The hole loses an energy
of $2J$ each time a p--h pair or a spinflip is excited,  its energy reaches
$\omega-6J$ after three emissions  and is only regained via three more
steps in which it absorbs the emitted pairs ({the spinflips are
immobile once created}). The maximal crossing of loops explains the order
the hole must follow in the last three steps, the only possible path at 6th
order is then the well--known Trugman loop (fig.\ref{fig:vertex6}c).
The delocalization of the hole is shown by
the momentum dependence of the renormalized selfenergy
(\ref{eq:vertex6}).

Obviously with closed paths of longer lengths the hole is
able to move further away from its initial position.
Evidently for any translation not affecting the spin structure it can
only end up on the same
sublattice, which in turn indicates that only vertex corrections of
order $4n+2$ will contribute. The next
nonzero contribution then appears naturally at 10th order.
Not supprisingly, the number of possible
paths grows considerably as one goes to higher orders.
fig.\ref{fig:vertex10} describes the three
processes responsible for the hole delocalization at 10th order.
Apparently, the first two processes share a common feature in that
in both cases the hole attempts to create a maximum number (here 5) of
spinflips before it eliminates them again. They differ
from one another only by the fact that they follow distinct paths in
eliminating the spinflips.
fig.\ref{fig:vertex10}a has maximal crossing of loops and is obviously
the direct extension of the 6th order correction. Its real space
configuration is therefore
described by fig.\ref{fig:real}a.
The hole follows a fixed path ({\it i.e.}, 1--2--3--4--5--6)
and hops
{\it two} loops minus {\it two} steps to its diagonal neighbor (i.e.
its first neighbor on the same sublattice) while erasing the
trace left behind. On the other hand, in fig.\ref{fig:vertex10}b the hole
changes its path starting from the 8th step and ends up at one of its
second nearest neighbors on the same sublattice (see fig.\ref{fig:real}b).
Finally in the third process (fig.\ref{fig:vertex10}c) we see at most
three spinflips in the intermediate states and the hole runs along the path
illustrated in fig.\ref{fig:real}c
and also moves to its second nearest neighbor sites.

To get an idea of how those higher order corrections affect the hole
mobility, at present we will limit ourselves to
the first category of closed paths. In other words, we shall only sum
the series of
maximally--crossed diagrams of $4n+2$ orders ($n=1,2,...$) as
demonstrated in fig.\ref{fig:max} to get a picture of the mobile hole
under Trugman's mechanism\cite{footnotes}. In a short while we shall
actually show that as far as low energy physics is concerned, the dominant
contribution comes only from the leading (6th) order vertex correction.
Fig.\ref{fig:max} is evaluated along the same lines
that lead to eq.(\ref{eq:vertex6}) and we get:
\begin{equation}
\Sigma^v({\bbox{k}},\omega+i0^+ )
=\frac{2}{N}\sum_{\bbox{q}}{\cal G}({\bbox{q}},\omega-2J)
\varepsilon_{\bbox{q}}\{
\varepsilon_{\bbox{q}}+V({\bbox{k}},{\bbox{q}})\} \;\;.
\label{eq:semax}
\end{equation}
We have introduced selfconsistency in order to simultaneously
account for the self--retracing paths discussed in the last section.
Thus, in the absence of $V({\bbox{k}},{\bbox{q}})$ this selfenergy just
reproduces the incoherent
result, eq.(\ref{eq:string}). $V({\bbox{k}},{\bbox{q}})$
sums the vertex corrections in fig.\ref{fig:max}. Using the bare Green
functions which are momentum--independent the momentum dependence of $V$ is
entirely due to the momentum dependence of the factors
$\varepsilon_{\bbox{k}}$ in eq.(\ref{eq:inter}). We then have
\begin{eqnarray}
V({\bbox{k}},{\bbox{q}})&=&\frac{2}{N}\sum_{{\bbox{q}}_2}
\frac{\varepsilon^2_{{\bbox{q}}_2}\varepsilon_{\bbox{k}+{\bbox{q}}_2
-{\bbox{q}}} (-t\varepsilon_{{\bbox{k}}+{\bbox{q}}_2})}
{(\omega-2J)(\omega-4J)^2(\omega-6J)}\nonumber\\
&+&\frac{2}{N}\sum_{{\bbox{q}}_2}
\frac{\varepsilon^2_{{\bbox{q}}_2}
\varepsilon_{{\bbox{k}}+{\bbox{q}}_2-{\bbox{q}}}
(-t\varepsilon_{{\bbox{k}}+{\bbox{q}}_2})^3}
{(\omega-2J)(\omega-4J)^2(\omega-6J)^2(\omega-8J)^2(\omega-10J)}+...... \}
\label{eq:vkq}
\;\;.
\end{eqnarray}
To help to understand this equation we note that ${\bbox{q}}_2$ in the above
expression
is the hole momentum  just one step before a complete closed path
(see the diagonal hole line in fig.\ref{fig:max}). In addition,
the only momentum dependence comes from the interaction vertex, and
summations over all but one internal momenta have been
performed noting that $-t\varepsilon_{\bbox{k}}$ is obtained by summing
${\bbox{q}}$ in $\varepsilon_{\bbox{q}}\varepsilon_{{\bbox{k}}-{\bbox{q}}}$.
The omitted higher order terms in (\ref{eq:vkq})
can be obtained straightforwardly remembering  that {\it bare} hole
lines and bubbles are used
inside the vertex part. The final result is determined by
solving the following equation:
\begin{equation}
V({\bbox{k}},{\bbox{q}})=-\frac{2z^2t}{N}
\sum_{{\bbox{k}}^\prime}
\frac{\gamma^2_{{\bbox{k}}^\prime-{\bbox{k}}}
\gamma_{{\bbox{k}}^\prime-{\bbox{q}}}}
{\gamma_{{\bbox{k}}^\prime}} F({\bbox{k}}^\prime,\omega) \;\;,
\end{equation}
\begin{equation}
F({\bbox{k}},\omega)=
\frac{(zt^2\gamma_{\bbox{k}})^2}
{(\omega-2J)(\omega-4J)^2(\omega-6J)}(F({\bbox{k}},\omega-4J)+1) \;\;.
\label{eq:iter}
\end{equation}
Without going into the details of a numerical solution,
we can immediately learn two points by examining the above
equation. First, as long
as $J/t\ll 1$ close to the bottom of the Brinkman--Rice spectrum
($\omega\sim\omega_0=-2\sqrt{z}t$) we have $F({\bbox{k}},\omega)
\approx
(zt^2\gamma_{\bbox{k}})^2/\omega^4$. In other words, the higher order vertex
corrections are quantitatively insignificant around the minimum of the
spectrum since every additional
two loops add a factor of roughly $1/16$. Therefore, the mobility of the
hole is dominantly determined by the leading vertex correction.
The second point needs some elaboration. The statement is that in any
event, the vertex correction above alone is not expected to give
quantitative satisfaction.
Since the point has in fact already shown up in eq.(\ref{eq:vertex6}),
we take the case of the leading vertex correction to demonstrate the
essential fact. Recall that in obtaining the preceeding equations
it is always implicitly assumed (by
using eq.(\ref{eq:bubble})) that the energy cost to create a spinflip
is $2J$. This is not correct in the presence of a hole.
{}From fig.\ref{fig:vertex6}c
we actually see that in fact it costs the hole only $3J/2$ to
perform the first step, $J$ for the next, and $J/2$ for the third
step, and vice versa for the energy--recovering steps. The evolution of
the energy cost for a hole moving around the plaquette is shown in
fig.\ref{fig:ener}a.
In the same manner, the $J$ factors in both eq.(\ref{eq:vkq}) and
eq.(\ref{eq:iter}) are seen not to give
a correct description of the energy cost in the intermediate states.
As can readily be understood this failure is directly related
to the mean field approximation we have applied in the
longitudinal channel (the third term in $H_1$): the
intraband correlations between holes at different intermediate
states have so far been neglected. The renormalizations due to
the correlation effects in this channel can be easily taken into account
at fixed order in the vertex corrections, as done in eq.(\ref{eq:sigmaf})
below. However, a systematic procedure to all orders seems to be difficult
to find because
the energy expense in the intermediate states of a closed path is
configuration--dependent. Take the 10th  order maximally crossed diagram
as example: we see the energy evolution shown in fig.\ref{fig:ener}b
which is apparantly unrelated to what we have observed in
fig.\ref{fig:ener}a. This situation is in sharp contrast with
the Brinkman--Rice case where
we know that except for the first step the energy cost
to extend the path by one step is always $J$.

After these remarks we now concentrate on the leading order vertex
correction. This can easily be done by truncating the
iteration eq.(\ref{eq:iter}) by its first order approximation.
In order to take into account the
correlations in the longitudinal channel we also include
renormaliztions so that the $J$ factors agree with the analysis of the
last paragraph (fig.\ref{fig:ener}a). Formally this can be achieved (at
fixed order in $H_I$) by using the second {\em and} the third term in
$H_1$ as the zeroth order Hamiltonian. The
final equation we then have is
\begin{equation}
\Sigma^v ({\bbox{k}},\omega+i0^+ )
=\frac{2z^2t^2}{N}\sum_{\bbox{q}}{\cal G}({\bbox{q}},\omega-3J/2) \{
\gamma^2_{\bbox{q}}
+\frac{2z^3t^4}{N R(\omega)}\sum_{{\bbox{q}}_2}
\gamma_{\bbox{q}}\gamma^2_{{\bbox{q}}_2}
\gamma_{{\bbox{k}}+{\bbox{q}}_2-{\bbox{q}}}
\gamma_{{\bbox{k}}+{\bbox{q}}_2}   \} \;\;,
\label{eq:sigmaf}
\end{equation}
where $R(\omega)=
(\omega-J/2)(\omega-3J/2)(\omega-5J/2)(\omega-3J)$.
To recover what we should have obtained in the absence of renormalizations,
we need only to replace $R(\omega)$ by
$(\omega-2J)(\omega-4J)^2(\omega-6J)$ and $(\omega-3J/2)$ in the
argument of $\cal G$ by $(\omega-2J)$.

We shall now solve the above equation numerically.
Fig.\ref{fig:density} shows our numerical results
for the density of states for $J=0.5t$ on a cluster of
20 sites. Since our calculation is reliable only near the bottom of the
spectrum and for relatively small $J/t$, one should note that
the data on the high
energy side of the spectrum does not make sense and is therefore omitted.
In the following the
discussion will be focused only on the lowest two peaks in the figure.
First of all our numerical solution shows that the hole
spectrum becomes dispersive with the
introduction of vertex corrections and the ground state is located at
${\bbox{k}}=(0,0)$. The density of states around the lowest peak
therefore broadens in contrast to the $\delta$--peak in the
Shraiman--Siggia limit. Secondly, the peaks
shift to lower energies when including the renormalization whereas the gap
between the lowest two peaks becomes narrower as is shown in
fig.\ref{fig:gap}a and can
readily be explained from the form of $R(\omega)$. Also remarkable
is the fact that
the gap fits nicely a law $\sim a J^{0.65}_z$ for $J_z\leq 0.4$, with or
without
the renormalizations (see the figure caption for the prefactor $a$).
This is close to the $J^{2/3}$--bahavior in the Shraiman--Siggia limit.
Let us now take a further look at the
quasiparticle peak at the bottom of the spectrum.
As was explained before, the nature of the quasiparticle
band does not depend too much upon the renormalizations we have
included as long as $J/t$ is small compared to unity.
In fig.\ref{fig:gap}b we show quasiparticle
residues for momenta at both the top and bottom of the band.
With or without the renormalizations, the two residues are remarkably
close to each other. The residue at the bottom of the
quasiparticle band is always the smallest for $J_z > 0.3$. Moreover,
$Z_{\bbox{k}=0}$ shows a $J^{0.86}_z$--bahavior for $J_z/t < 0.6$.
These results
agree well with numerical findings by exact
diagonalization\cite{poilblanc_scaling_1hole}.
In fact, the structure of the quasiparticle band can actually be obtained
analytically now that only the leading vertex correction are included.
With the dominant pole approximation\cite{kane_slave_1hole}, we readily
find that the quasiparticle band obeys:
\begin{equation}
\varepsilon_{\bbox{k}}=\varepsilon_0- \frac{J}{4z^2}\{(\cos
k_x+\cos k_y)^2 + \cos (k_x+k_y)+\cos (k_x-k_y)\} \;\;,
\label{eq:ekt}
\end{equation}
where $\varepsilon_0$ is a constant. The quasiparticle is therefore seen to
be  confined within a narrow band of bandwidth $J/z^2$ and
the minimum  at ${\bbox{k}}=(0,0)$.
Finally, the ground state energy for the hole exhibits the
$J_z^{2/3}$--behavior, simply following the string scenario. This should
not be a surprise in view of the energy scale
of the vertex correction near the bottom of the band.

\subsection{The coherent spectrum: spin fluctuations}
Finally we switch back on the
spin fluctuations which necessarily favor the delocalization of
the hole. For simplicity, we shall take into account only the fluctuations
in the transverse spin channel. The interaction part in $H_1$ is
then composed of eq.(\ref{eq:inter}) and the
the Heisenberg exchange in the transverse direction. As a immediate
consequence we now have to consider also the process described by
fig.\ref{fig:2nd}b since
the hole may easily face a ``wrong--spin'' neighbor created through
spin fluctuations. It is certainly likely to hop to that
neighbor in order to restore the local N\'eel ordering by absorbing the
overturned spin. On the other hand, if the hole does hop to one of its
N\'eel ordered neighbors, spin fluctuations will help to lower the
energy by dispersing the newly created spinflip through the medium.
In summary, spin fluctuations
facilitate the propagation of the hole either by creating a trap
in front of it or by immediately eliminating the trace it leaves
behind. In this section we shall limit ourselves only to this
new mechanism, while neglecting  Trugman's mechanism assuming
that the latter plays only a relatively minor role. This neglect is in fact
justified by the numerically small prefactor in front of
the $\bbox{k}$--dependent term
in (\ref{eq:ekt}) ($J/4z^2 = J/64$).

In order to understand the diagrammatic description, let us look at the
case in which a downspin hole initially sits on the origin and is
surrounded by a local
N\'eel environment. As it hops to one of its neighbors, it turns to an
upspin hole at the new position
and creates a spinflip at the origin. The flipped spin
will now propagate in the medium because of spin fluctuations. Suppose that
it approaches the hole again  and ends up at one of its
neighbors (on the even sublattice). For the reasons discussed above the hole
would like to eliminate that overturned spin. This can happen in two
different ways.
It may either directly hop there, or it may hop one step more to
another neighbor and let the newly created spinflip be eliminated
together with the previous one through
spin fluctuations. In the first case, all it involves in the
intermediate states
is the propagation of a spinflip from the origin to a site on the same even
sublattice. The Feynman diagram for this process is therefore
fig.\ref{fig:flu}a, where the propagation of the intermediate spinflip
is given by $\chi_{11}({\bbox{q}},\omega)$
from eq.(\ref{eq:mode}). The selfenergy of the hole is then determined by
\begin{equation}
\Sigma^a_\downarrow({\bbox{k}},\omega+i0^+)
=i\frac{2z^2t^2}{N}\sum_{{\bbox{q}}}\int\frac{d\omega_1}{2\pi}
{\cal G}_\uparrow^v({\bbox{k}}-{\bbox{q}},\omega-\omega_1+i0^+)
\gamma^2_{{\bbox{k}}-{\bbox{q}}}\chi_{11}({\bbox{q}},\omega_1) \;\;.
\label{selfa}
 \end{equation}
It differs from eq.(\ref{eq:se1}) only by the fact that
the intermediate particle--hole pair is now able to propagate.
On the other hand, in the second case we see an even number of
particle--hole
bubbles in the intermediate state between two sucessive hops of the
hole. The process is therefore  described by fig.\ref{fig:flu}b and
we obtain:
\begin{equation}
\Sigma^b_\downarrow({\bbox{k}},\omega+i0^+)
=i\frac{2z^2t^2}{N}\sum_{{\bbox{q}}}\int\frac{d\omega_1}{2\pi}
{\cal G}_\uparrow^v({\bbox{k}}-{\bbox{q}},\omega-\omega_1+i0^+)
\gamma_{{\bbox{k}}-{\bbox{q}}}\gamma_{\bbox{k}}\chi_{21}({\bbox{q}},\omega_1)
\;\;. \label{selfb}
\end{equation}
Some attention should be paid to the interaction vertex (namely
$\gamma_{\bbox{k}}$) at the second
hoping process. Its momentum dependence is determined by that of the hole in
the final state ({\it i.e.} ${\bbox{k}}$). As for the dynamical
susceptibility,
it is given by the off--diagonal term in eq.(\ref{eq:mode}) since we
actually see the propagation of a pair of spinflips created by two
successive hopings. An analogous analysis applies if
the hole is initially surrounded by distorted spins. The corresponding
Feynman diagrams are shown in fig.\ref{fig:flu}c and
fig.\ref{fig:flu}d, from which we find:
\begin{equation}
\Sigma^c_\downarrow({\bbox{k}},\omega+i0^+)
=i\frac{2z^2t^2}{N}\sum_{{\bbox{q}}}\int\frac{d\omega_1}{2\pi}
{\cal G}_\uparrow^v({\bbox{k}}-{\bbox{q}},\omega-\omega_1+i0^+)
\gamma^2_{\bbox{k}}\chi_{22}({\bbox{q}},\omega_1) \;\;,
\label{selfc}
\end{equation}
and
\begin{equation}
\Sigma^d_\downarrow({\bbox{k}},\omega+i0^+)
=i\frac{2z^2t^2}{N}\sum_{{\bbox{q}}}\int\frac{d\omega_1}{2\pi}
{\cal G}_\uparrow^v({\bbox{k}}-{\bbox{q}},\omega-\omega_1+i0^+)
\gamma_{\bbox{k}}\gamma_{{\bbox{k}}-{\bbox{q}}}\chi_{12}({\bbox{q}},\omega_1)
\;\;, \label{selfd}
\end{equation}
respectively. Finally we integrate over frequency and sum
eqs.(\ref{selfa})--(\ref{selfd}) to get the total selfenergy:
\begin{equation}
\Sigma_\downarrow({\bbox{k}},\omega)=
\frac{2}{N}\sum_{\bbox{q}} f({\bbox{k}},{\bbox{q}})
{\cal G}_\uparrow^v({\bbox{k}}-{\bbox{q}},\omega-\omega_{\bbox{q}}) \;\;,
\label{eq:selast}
\end{equation}
\begin{equation}
f({\bbox{k}},{\bbox{q}})=\frac{1}{2}z^2t^2\{
(\frac{1}{\sqrt{1-\gamma^2_{\bbox{q}}}}+1)\gamma^2_{{\bbox{k}}-{\bbox{q}}}+
(\frac{1}{\sqrt{1-\gamma^2_{\bbox{q}}}}-1)\gamma^2_{\bbox{k}}-
\frac{2\gamma_{{\bbox{k}}-{\bbox{q}}}\gamma_{\bbox{k}}
\gamma_{\bbox{q}}}{\sqrt{1-\gamma^2_{\bbox{q}}}}\} \;\;.
\label{eq:Vdressed}
\end{equation}

The  three terms appearing in the dressed vertex function
$f({\bbox{k}},{\bbox{q}})$
reflect the different scattering processes we have justed discussed.
(The factor 2 in the last term comes from the degeneracy in the
processes leading to  $\Sigma^b$ and $\Sigma^d$). In particular
$f({\bbox{k}},{\bbox{q}})$ shows unambiguously that
for a hole on the ``nesting line'' $|k_x|+|k_y|=\pi$ (where
$\gamma_{\bbox{k}}= 0$),
the only nonzero contribution comes from $\Sigma^a$. The hole having these
particular momenta  responds
{\it only} to its N\'eel ordered neighbors. In addition, this hole only
couples weakly to the dispersing low energy spin excitations, leading to
$f({\bbox{k}},{\bbox{q}})\sim
({\bbox{q}}_x+{\bbox{q}}_y)^2/|{\bbox{q}}|$, which in turn translates into
$Im\Sigma\sim\omega^2$ in two dimensions\cite{kane_slave_1hole}.
In other words, quasiparticle behavior is
well defined for these particular momenta.
Secondly, quantum fluctuations play their full role in the hole
dynamics only away from nesting wavevectors.
Nonetheless, at these momenta none of the
aforementioned scatterings by themselves
support quasiparticle picture, as is clearly indicated by the
individually divergent terms in (\ref{eq:Vdressed}).
Only from the sum of all the terms
do we obtain $f({\bbox{k}},{\bbox{q}})\sim |{\bbox{q}}|$ or
$Im\Sigma\sim\omega^2$ in the long
wavelength limit and hence a quasiparticle behavior  for the single hole.
Finally, we notice that eqs.(\ref{eq:selast},\ref{eq:Vdressed})
are exactly the same as those
obtained by Schmitt--Rink et~al.\cite{schmittrink_slave_1hole} and by Kane
et~al.\cite{kane_slave_1hole} using the slave--fermion technique.
Numerical calculations based on eqs.(\ref{eq:selast}) and (\ref{eq:Vdressed})
have been carried out by several
groups\cite {marsiglio_slave_1hole,martinez_slave_1hole,liu_slave_1hole}
and they have indeed observed a well--behaved quasiparticle peak
near the bottom of the spectrum. In particular, the ground
state is found to be at ${\bbox{k}}=(\pi/2,\pi/2)$ where, as we now see,
a quasiparticle picture is well estabilished.
In ref.\onlinecite{poilblanc_scaling_1hole} the numerical results obtained by
the
above groups were also compared with  exact diagonalization results. Good
agreement has been  found concerning the ${\bbox{k}}$ dependence of
the quasiparticle residue as well as its variation with $J$.

Before ending this section, we remind the reader that
in studying the dynamics of a single hole we have assumed from the start
the scattering processes involving energies of $O(U)$ are irrelevant
as far as low energy properties of the single hole are concerned. To see
exactly how this comes about we can in principle follow the same reasoning as
for half filling. In the presence of spin fluctuations the
selfconsistent RPA is then formed within the entire Hilbert space
including scatterings involving double occupancies. The corresponding
Feynman diagrams are those of fig.\ref{fig:flu} except that the
bare bubbles are replaced by those of fig.\ref{fig:rpa}b. The final result is
therefore again eq.(\ref{eq:selast}) and eq.(\ref{eq:Vdressed}).
In concluding this section, we also cite the recent calculation performed
by Brenig and Kampf\cite{brenigkampf_1hole}. These authors have studied
the dynamics of a single hole in the presence of a SDW antiferromagnet
which is then extrapolated  to the large--U limit.
When a selfconsistent RPA scheme is adopted (as is done here), they have
observed, along with the incoherent background, a quasiparticle band with a
significantly reduced bandwidth. Moreover, the quasiparticle residue is
found maximal at ${\bbox{k}}=(\pi/2,\pi/2)$ and minimal at
${\bbox{k}}=(0,0)$. These observations agree with those derived from
eq.(\ref{eq:selast}) and eq.(\ref{eq:Vdressed}).

\section{Summary}
We summarize our main results in this section.
We have shown that a transformation to a new basis
facilitates the investigation of the crossover
between weak and strong coupling limits of the Hubbard model.
The new basis has been introduced via a Bogolyubov rotation.
Its relation with the weak coupling method is seen from
the fact that the transformation had previously been applied
to the Hartree--Fock
Hamiltonian
which in turn served as the basis for further studies of
fluctuation effects.
Indeed,
for the case of half filling, we have
explicitly demonstrated in section \ref{sec:half} the smooth crossover
between two limits of Hubbard correlations.
High energy scatterings have been shown to be irrelevant and are
scaled away as U approaches the strong coupling regime.
The correct spin wave mode we have derived in the final result is actuallly
 rather evident since the exact Heisenberg interaction in the transverse
spin channel
is contained in the low energy part of the effective Hamiltonian ({\it
i.e.} eq.(\ref{eq:Heff})).

Within the new basis, we have further
studied in detail the dynamics of a single hole in the
antiferromagnetic background. Aside from the fact that we can
diagrammatically distinguish the scatterings responsible for the low energy
physics from those at high energy, we are also able to
separate in Feynman diagrams the incoherent from coherent physical
processes.
In the Ising limit, we have analyzed a series of vertex corrections
responsible for
the hole delocalization and have concluded that the
leading vertex correction dominates as far as low energy physics is
concerned.
 In particular, for small $J/T$ quantitative agreement is reached with exact
calculations on small clusters regarding the
low energy physics, namely the ground state and the nature
of the coherent motion for a single hole near the bottom of the spectrum.
The transverse spin fluctuations further support a quasiparticle
picture in the case of a single hole. We have derived
a selfconsistent equation  which recovers previous results
for the $t-J$ model which were obtained using slave--fermion methods.
The numerical solution based on that selfconsistent equation
also shows good agreement with exact solutions on small clusters.
Finally, agreement is also found with results derived from
direct extrapolation of weak coupling calculations recently done by
Brenig and Kampf\cite{brenigkampf_1hole}.

An important ingredient in the present calculation
is the two
sublattice antiferromagnetic spin structure. This is certainly correct
in the case of
one hole. However, finite
hole--doping will certainly destroy the antiferromagnetic
spin structure on which our discussions are based.
The question concerning the stability of the
quasiparticle
picture and in particular its relevance to physical properties at finite
hole densities therefore remains open. Further, the change of physical
properties as one goes from weak to strong correlation at finite doping has
also not been fully clarified.
We hope to address this question in a forthcoming investigation.

\newpage
\appendix
\section{}
In this appendix we explain the procedure which leads to eq.(\ref{eq:hred}),
i.e. the Hubbard Hamiltonian written within the reduced Brillouin zone (RBZ).

By definition, the Fourier transform of a fermion operator at site $j$
is expressed as
\begin{equation}
C_j=\frac{1}{\sqrt{N}}{\sum_{\bbox{k}}}^\prime e^{i{\bbox{k}}\cdot{\bf j}}
(C_{\bbox{k}}+(-1)^jC_{{\bbox{k}}+{\bbox{Q}}}) \;\;,
\end{equation}
where the summation over ${\bbox{k}}$ covers the RBZ (i.e.
$|k_x|+|k_y|\le \pi$) and ${\bbox{Q}}=(\pm\pi,\pm\pi)$.
$C_{{\bbox{k}}+{\bbox{Q}}}$ complements $C_{\bbox{k}}$ and together they
run over every
state in the whole Brilliuon zone once and only once. Correspondingly,
the number operator is written as
\begin{equation}
n_j=\frac{1}{N}{\sum_{{\bbox{q}}{\bbox{k}}}}^\prime e^{-i{\bbox{q}}\cdot{\bf
j}}
(C^\dagger_{{\bbox{k}}+{\bbox{q}}}+
(-1)^jC^\dagger_{{\bbox{k}}+{\bbox{q}}+{\bbox{Q}}})
(C_{\bbox{k}}+(-1)^jC_{{\bbox{k}}+{\bbox{Q}}}) \;\;,
\end{equation}
where again the summations for both ${\bbox{q}}$ and ${\bbox{k}}$ are
confined to within the RBZ. In the meantime,
$C_{{\bbox{k}}+{\bbox{q}}}$ and $C_{{\bbox{k}}+{\bbox{q}}+{\bbox{Q}}}$ are
complementary and cover the whole Brillouin zone centered at
${\bbox{q}}$. The above expression can then be written compactly as
\begin{equation}
n_j=\frac{1}{N}{\sum_{{\bbox{q}}{\bbox{k}}}}^\prime e^{-i{\bbox{q}}\cdot{\bf
j}}
\{\Psi^\dagger({\bbox{k}}+{\bbox{q}})
\Psi({\bbox{k}})+(-1)^j\Psi^\dagger({\bbox{k}}+{\bbox{q}})
\sigma^1\Psi({\bbox{k}})\}
\end{equation}
by defining a spinor $\Psi^\dagger_\sigma
({\bbox{k}})=(C^\dagger_{{\bbox{k}}\sigma},
C^\dagger_{{\bbox{k}}+{\bbox{Q}}\sigma})$.
The Hamiltonian can then readily be transformed noting that
the following equality holds for the Fourier transform on sublattices:
\begin{equation}
\frac{2}{N}{\sum_r}^\prime e^{-i{\bbox{q}}\cdot{\bf r}}=\delta_{{\bbox{q}},0}
\;\;,
\end{equation}
where the summation runs on either sublattice and ${\bbox{q}}$ is defined
within the
RBZ. The $U$--term in the Hubbard model can then be expanded as
\begin{eqnarray}
\nonumber
U\sum_i n_{i\uparrow}n_{i\downarrow} & = &
\frac{U}{N}\sum_{{\bbox{k}}{\bbox{k}}\prime{\bbox{q}}}\{\Psi^\dagger_\uparrow
({\bbox{k}}+{\bbox{q}})\Psi_\uparrow ({\bbox{k}})\Psi^\dagger_\downarrow
({\bbox{k}}^\prime-{\bbox{q}})
\Psi_\downarrow ({\bbox{k}}^\prime) \\
& & \quad \quad +
\Psi^\dagger_\uparrow({\bbox{k}}+{\bbox{q}})\sigma^1\Psi_\uparrow
({\bbox{k}})
\Psi^\dagger_\downarrow
({\bbox{k}}^\prime -{\bbox{q}})\sigma^1\Psi_\downarrow ({\bbox{k}}^\prime)\}
\;\;.
\end{eqnarray}
Eq.(\ref{eq:hred}) is thus derived.

\section{}
In this appendix we detail the procedure which transforms the
susceptibility matrix, eq.(\ref{eq:matrixA}), so that it is represented by
the $r$--operators, i.e. in terms of Wannier functions.
Following appendix A, the spin operator
$\bar{S^+_j}=C^\dagger_{j\uparrow}
C_{j\downarrow}$ at site $j$ can be written within the RBZ as
\begin{equation}
S ^+_j=\frac{1}{N}{\sum_{{\bbox{q}}{\bbox{k}}}}^\prime
e^{i{\bbox{q}}\cdot{\bf j}} \{\Psi^\dagger_\uparrow({\bbox{k}}-{\bbox{q}})
\Psi_\downarrow({\bbox{k}})+(-1)^j
\Psi^\dagger_\uparrow({\bbox{k}}-{\bbox{q}})\sigma^1
\Psi_\downarrow({\bbox{k}})\} \;\;.
\end{equation}
One can now Fourier transform
the susceptibility matrix given in eq.(\ref{eq:matrixA})
within the RBZ and then write it in terms of $r$--operators by
inverting eq.(\ref{eq:rotation}). The final result for half filling
is
\begin{equation}
\chi^{-+} ({\bf x},{\bf x}^\prime,t) = \frac{2}{N}
\sum_{\bbox{q}} \chi^{-+}({\bbox{q}},t) e^{ i{\bbox{q}}\cdot({\bf x}-{\bf
x}^\prime) } \;\;,
\end{equation}
\begin{eqnarray}
\chi^{-+}({\bbox{q}},t) = -\frac{i}{2N} \sum_{{\bbox{k}}{\bbox{k}}^\prime}
&  & \left(
f_1({\bbox{k}})f_1({\bbox{k}}^\prime)
\left(\begin{array}{cc}a_{11}&a_{12}\\a_{21}&a_{22}\end{array}\right)
+f_1({\bbox{k}})f_2({\bbox{k}}^\prime)
\left(\begin{array}{cc}a_{12}&a_{11}\\a_{22}&a_{21}\end{array}\right)
\right.
\nonumber\\
 & + &\left. f_2({\bbox{k}})f_1({\bbox{k}}^\prime)
\left(\begin{array}{cc}a_{21}&a_{22}\\a_{11}&a_{12}\end{array}\right)
+f_2({\bbox{k}})f_2({\bbox{k}}^\prime)
\left(\begin{array}{cc}a_{22}&a_{21}\\a_{12}&a_{11}\end{array}\right)
\right) \label{eq:matrixB}
\;\;,
\end{eqnarray}
where
$f_1 ({\bbox{k}})=
(u_{\bbox{k}}+v_{\bbox{k}})(u_{{\bbox{k}}-{\bbox{q}}}
+v_{{\bbox{k}}-{\bbox{q}}})$, $f_2 ({\bbox{k}})=
-(u_{\bbox{k}}-v_{\bbox{k}})(u_{{\bbox{k}}-{\bbox{q}}}-
v_{{\bbox{k}}-{\bbox{q}}})$, and \begin{eqnarray}
\nonumber
\lefteqn{%
\left(\begin{array}{rr} a_{11}&a_{12}\\ a_{21}&a_{22}\end{array}\right)
=} \\
& & \langle T
\left(\begin{array}{rr}
r^{v \dagger}_{{\bbox{k}}\downarrow}(t)r^c_{{\bbox{k}}-{\bbox{q}}\uparrow}(t)
r^{c \dagger}_{{\bbox{k}}^\prime-{\bbox{q}}\uparrow}(0)
r^v_{{\bbox{k}}\prime\downarrow}(0)&
r^{v \dagger}_{{\bbox{k}}\downarrow}(t)r^c_{{\bbox{k}}-{\bbox{q}}\uparrow}(t)
r^{v \dagger}_{{\bbox{k}}^\prime-{\bbox{q}}\uparrow}(0)
r^c_{{\bbox{k}}\prime\downarrow}(0)\\
r^{c \dagger}_{{\bbox{k}}\downarrow}(t)r^v_{{\bbox{k}}-{\bbox{q}}\uparrow}(t)
r^{c \dagger}_{{\bbox{k}}^\prime-{\bbox{q}}\uparrow}(0)
r^v_{{\bbox{k}}\prime\downarrow}(0)&
r^{c \dagger}_{{\bbox{k}}\downarrow}(t)r^v_{{\bbox{k}}-{\bbox{q}}\uparrow}(t)
r^{v \dagger}_{{\bbox{k}}^\prime-{\bbox{q}}\uparrow}(0)
r^c_{{\bbox{k}}\prime\downarrow}(0)
\end{array}\right)\rangle \;\;.
\end{eqnarray}
As shown in eq.(\ref{eq:matrixB}), we need in general to calculate response
functions
with momentum dependent vertex functions $f_1({\bbox{k}})$ or
$f_2({\bbox{k}})$ attached to either of the two external points.
Nevertheless, for $U\gg t$, we can easily verify that
$\chi^{-+}$ is dominated by the first term in eq.(\ref{eq:matrixB}).
The remaining terms are smaller by at least a factor $O(J/U)$.

\newpage

\begin{thebibliography}{10}
\bibitem{footnotes3}
More precisely, we have
$\chi^{-+}({\bbox{q}},\omega)=-\chi^{-+}_{ST}({\bbox{q}} ,-\omega)$.
  The minus sign in front comes from the definition, while the minus
 sign before $\omega$ comes from the assumption we make in this paper
that the spin points
  down on the even sublattice. The comparison with
  SWZ (ref.\onlinecite{schrieffer_spinbag})
can be done by using the relations given in
  ref.\onlinecite{footnotes2}.
\bibitem{footnotes}
For maximally--crossed diagrams of both 6th and 10th orders
the final state of the hole hops is its diagonal nearest neighbor.
This can not be generalized however. As can be best demonstrated
at 14th order, higher order maximally--crossed diagrams
also allow the hole to hop beyond its diagonal neighbors.
\end{thebibliography}

\newpage

\begin{figure}\caption{  (a) The RPA scheme for the spin--spin correlation
       function ($\chi_{11}$) in the transverse spin
       channel. The dotted
       lines between bubbles denote the $g_2$ interactions.
  The ``bare'' bubble here is the result of (b).
       (b) The interband multiple scatterings between a $c$--particle and
       $v$--hole
       which result in the renormalized ``bare'' bubble used in (a). The
       interaction vertex in the ladder are $g_1$.
\label{fig:rpa}}\end{figure}

\begin{figure}\caption{The 1st order nonzero contribution to $\chi_{21}$ in
       eq.(\protect\ref{eq:matrix}). The ``bare'' bubble in the figure
       is the $g_1$--renormalized result given in fig.\protect\ref{fig:rpa}b.
       As in Fig.\protect\ref{fig:rpa} the dotted lines denotes the
       $g_2$ interactions.
\label{fig:rpa2}}\end{figure}

\begin{figure}\caption{The second order correction to the hole selfenergy.
       For a downspin
       hole, (a) is determined by the first term in $H_I$, while (b) by the
       second term. (b) does not contribute in the absence of
       spin fluctuations (see text).
       The dashed line joining the bubble and the hole line
       denotes the $O(t)$ interactions in $H_I$.
\label{fig:2nd}}\end{figure}

\begin{figure}\caption{The selfconsistent  2nd--order correction to the
       selfenergy which accounts for the incoherent background
       in the hole spectrum.
\label{fig:self2nd}}\end{figure}
\begin{figure}\caption{A typical selfenergy diagram in the
       Shraiman--Siggia limit.
\label{fig:SS}}\end{figure}

\begin{figure}\caption{  The leading vertex correction in the Ising limit.
       (b) is an equivalent of (a), in which the particle--hole bubble
is denoted
       by a dotted line; (c) The real space configuration corresponding to
 (a) or (b).
\label{fig:vertex6}}\end{figure}
\begin{figure}\caption{  Nonzero vertex corrections at 10th order.
       The dotted line denotes a particle--hole bubble of opposite spins as in
       fig.\protect\ref{fig:vertex6}. Note that the spin of
the $c$--particle in each
       bubble is the same as that of the intermediate third hole.
\label{fig:vertex10}}\end{figure}

\begin{figure}\caption{Schematical representation in real space of the
graphs in fig.\protect\ref{fig:vertex10}.
\label{fig:real}}\end{figure}

\begin{figure}\caption{ Maximally--crossed diagrams representing the
       coherent motion of a single hole in the Ising limit.
       Fig.\protect\ref{fig:vertex6}b and fig.\protect\ref{fig:vertex10}a
       are redrawn here in a different fashion to guide the eye.
\label{fig:max}}\end{figure}

\begin{figure}\caption{The evolution of the energy in the intermediate
states immediately after
the $n$th hopping step: (a) for the leading vertex correction; (b)
	for the 10th order maximally crossing diagram (see
	fig.\protect\ref{fig:vertex10}a).
\label{fig:ener}}\end{figure}

\begin{figure}\caption{The density of
	states for a single hole in the Ising limit at $J/t=0.5$,
calculated on the 20--site cluster.
A Lorentzian broadening with broadening parameter
	$\delta=0.01$ is used. (a) With renormalzations as explained in the
	text, (b) without renormalizations.
\label{fig:density}}\end{figure}

\begin{figure}\caption{(a) The gap $\Delta$ between the lowest two states
as a finction of $J_z$ (Ising limit),
	 with renormalization (lower curve) and without renormalzation
	 (upper curve) due to interactions in the longitudinal channel.
	 The lines are fits to $\Delta=2.610J_z^{0.65}$ and
	 $\Delta=3.173J_z^{0.65}$, respectively.
	 (b) The quasiparticle residue for ${\protect\bbox{k}}=(0,0)$,
	 denoted by ( + ) and ${\protect\bbox{k}}=(0,\pi)$, denoted
	 by ( * ), with renormalization (lower curves) and without
	 renormalzation (upper curves). The lines are fits to
	 $Z_{{\protect\bbox{k}=0}}=0.511J^{0.86}_z$ and
	 $Z_{{\protect\bbox{k}=0}}=0.662J^{0.86}_z$, respectively.
\label{fig:gap}}\end{figure}

\begin{figure}\caption{The selfenergy diagram in the presence
of spin fluctuations:
	(a) the hole creates and then eliminates a spinflip at two
	successive hops; (b) it creates a spinflip at each of the two
	hops; (c) it first eliminates and then create a spinflip;
	(d) it eliminates a spinflip at each of the two hops.
	In (a) and (b) the hole faces a N\'eel surrounding in the initial
	state, while in (c) and (d) it sees one of its neighbors disturbed
	in the initial state.
\label{fig:flu}}\end{figure}

\end{document}